# Estimating Road Network Accessibility during a Hurricane Evacuation: A Case Study of Hurricane Irma in Florida[1]

Yi-Jie Zhu[1], Yujie Hu[2,3*], Jennifer M. Collins[1]

[1]School of Geosciences, University of South Florida, Tampa, FL 33620, USA
[2]Department of Geography, University of Florida, Gainesville, FL 32611, USA
[3]UF Informatics Institute, University of Florida, Gainesville, FL 32611, USA

*Corresponding author: Yujie Hu
Email: yujiehu@ufl.edu

**Abstract**

Understanding the spatio-temporal road network accessibility during a hurricane evacuation—the level of ease of residents in an area in reaching evacuation destination sites through the road network—is a critical component of emergency management. While many studies have attempted to measure road accessibility (either in the scope of evacuation or beyond), few have considered both dynamic evacuation demand and characteristics of a hurricane. This study proposes a methodological framework to achieve this goal. In an interval of every six hours, the method first estimates the evacuation demand in terms of number of vehicles per household in each county subdivision by considering the hurricane's wind radius and track. The closest facility analysis is then employed to model evacuees' route choices towards the predefined evacuation destinations. The potential crowdedness index ($PCI$), a metric capturing the level of crowdedness of each road segment, is then computed by coupling the estimated evacuation demand and route choices. Finally, the road accessibility of each sub-county is measured by calculating the reciprocal of the sum of $PCI$ values of corresponding roads connecting evacuees from the sub-county to the designated destinations. The method is applied to the entire state of Florida during Hurricane Irma in September 2017. Results show that I-75 and I-95 northbound have a high level of congestion, and sub-counties along the northbound I-95 suffer from the worst road accessibility. In addition, this research performs a sensitivity analysis for examining the impacts of different choices of behavioral response curves on accessibility results.

---

## 1. Introduction

Hurricanes are devastating natural hazards due to their strong winds, torrential rains, inland flooding of water-bodies, and coastal flooding by storm surges (Powell & Reinhold, 2007; Rezapour & Baldock, 2014). During past decades, the United States has experienced several major hits and impacts from hurricanes. This includes the unforgettable hurricane season in 2017 when damage costs and life losses were historically high. The trio of hurricanes (Harvey, Irma, Maria) imposed nearly 370 billion dollars of loss to the United States and its territories (Halverson, 2018). Particularly in the Hurricane Irma case, a large-scale evacuation order was made across the State of Florida to avoid severe life losses. In terms of hurricane evacuation in the United States, state governments designate specific routes—usually highways—for evacuation and suggest departure time for residents to evacuate from a hurricane threat area. However, in practice, studies on evacuation behaviors indicate the existence of a synchronized phenomenon when most evacuees not only delay their departure but also choose similar routes based on their own risk perceptions (Yang et al., 2016). This behavior brings about traffic surges and potential congestion in the road network, causing unpredicted evacuation delays and thus potentially putting more lives at risk. Such elevated evacuation traffic could become much worse during a large-scale evacuation in Florida, such as the Irma case. Specifically, because of Irma's track, Florida's peninsula shape, and the whole of the peninsula at times falling within the cone of uncertainty as forecasted by the National Hurricane Center, most evacuees in Florida had no choice but to travel northwards, causing extreme evacuation traffic delay in the northbound lanes of several interstate highways such as I-75 linking Florida and Georgia. According to the officials, a 1236% traffic increase on northbound I-75 was recorded (Florida Department of Transportation, 2018). Therefore, modeling the road network accessibility during a hurricane evacuation, or the level of ease of residents in an area in reaching evacuation destinations through the road network, is of great importance to evacuation management and planning (Tayler, 2012; Üster et al., 2018).

Existing research in assessing road accessibility—both evacuation-related and beyond—can be summarized into two main types of approaches—*topological-based* and *system-based*. The topological-based approach is built on an abstract node-link model, whereas the system-based approach overcomes some limitations from the topological-based approaches (e.g. missing dynamic effects of network disruption) by integrating more information regarding dynamic traffic demand and supply (Mattsson & Jenelius, 2015). Both methods have been extensively applied in previous studies on road network accessibility analysis (e.g. Duan & Lu, 2014; Sullivan et al., 2010). The topological-based approach was broadly used in non-specified hazard scenarios because it requires less data input (Dehghani et al., 2014). For example, Duan & Lu (2014) used this method to simulate hypothetical attacks in six cities by successively removing nodes from the abstract network. In contrast, the system-based approach is more widely used in analyses concerning specified natural hazard scenarios such as flooding and earthquakes (Suarez et al., 2005; Sohn, 2006; Khademi et al., 2015).

While many measures and approaches have been proposed for studying road network accessibility during a hazard, few studies have focused on evacuation from a hurricane, a particular hazard for coastal areas, as well as other communities. Different from other non-notice hazards that require immediate evacuation (e.g. an earthquake or terrorist attack), evacuation by hurricanes is dependent on the predicted track and intensity of the storm as well as people's perceptions of risk. Many factors can affect road network accessibility, and two of the most important ones are road carrying capacity and travel demand. As the road carrying capacity is usually fixed unless undergoing expansion or counter-flow usage, travel demand is, therefore, the main determinant of road accessibility during a hurricane evacuation. Within this particular scope of research, most existing studies extensively rely on collecting post-hurricane surveys for estimating evacuation travel demand. For example, in the context of Hurricane Bonnie in 1998, North Carolina, Whitehead et al. (2000) examined the determinants of evacuation behavior based on a telephone survey of coastal residents. Among all the factors they assessed, the risk factor (wind and flood) is reported to be the most important predictor of hurricane evacuation compared to socioeconomic factors.

To better understand road accessibility levels during a hurricane evacuation, Dow and Cutter (2002) held a survey of coastal South Carolina residents about their decisions during the 1999 Hurricane Floyd evacuation. The study shows that a majority of evacuees decided to use interstate highways, such as I-26, despite the heavy traffic congestion. The result reflects the evacuees' route choice preferences and suggests interstate highways as the main target for future road accessibility studies on hurricane evacuation. The aforementioned survey-based research helps to understand evacuation demand but lacks quantitative linkages to the properties of the hurricane. Wilmot and Meduri (2005) developed a method to assess evacuation zones for estimating evacuation demand. In their model, properties of storm surges were considered. However, this model may only be applicable to coastal areas where storm surge induced flooding is a major concern, and thus may not be directly applied to cases where a massive evacuation including inland residents is ordered. Another way to recognize the influence of a hurricane on evacuation demand and behavior is to consider the distance to the storm center from where residents live (Fu and Wilmot, 2006). The probability of residents in an area choosing to evacuate increases when the storm (center) approaches the area. This is described by the behavioral response curves which represent different departure time choices made by the evacuees (Fu et al., 2007). The response curves are commonly modeled by distance decay functions of varying types. Simply setting a distance decay function from residents' homes to the storm center may lead to biased estimations of evacuation demand, however. This is because the impact of a hurricane can spread hundreds of kilometers away from the core. Therefore, they set a distance threshold to the storm core within which residents would not consider evacuating. Some other studies (e.g. Yuan, et al., 2006; Lindell, 2008) used empirical data to estimate evacuation demand based on different behavioral response curves. The results of these studies vary case by case due to different characteristics of the hurricane and inconsistent geographies under investigation. Therefore, a more systematic way to measure road network accessibility that can account for both evacuation demand and physical factors of a hurricane, such as its wind speed and track, is imminently needed.

This study provides a methodological framework for evaluating road accessibility during a hurricane evacuation by considering both dynamic evacuation demand and hurricane characteristics. It first estimates the evacuation demand in terms of the number of vehicles per household in each county subdivision (i.e., sub-county) by considering the hurricane's wind radius and track. The closest facility analysis in Geographic Information Systems (GIS) is then employed to model evacuees' route choices towards the predefined evacuation destinations. Next, a metric termed the potential crowdedness index ($PCI$), which is defined to capture the crowdedness or congestion level of each road segment, is computed by coupling the estimated evacuation demand and route choices. Finally, the road accessibility of each sub-county is obtained by calculating the reciprocal of the sum of $PCI$ values of corresponding roads connecting evacuees from the sub-county to the designated destinations. To illustrate this method, it is applied to the case of Hurricane Irma (September 2017) where the cone of uncertainty covered the entire peninsula of Florida, and with a track in a general north direction through the state, triggering the majority of those who evacuated to head towards the state's northern boundary (Wong et al, 2018). The results and spatial maps can help identify low-access road segments and highlight the vulnerable neighborhoods, and hence can shed light on evacuation management and transportation planning.

**2. Study Area and Data Sources**

The state of Florida in the United States, surrounded on three sides by subtropical waters, is one of the most vulnerable states to hurricane hazards due to its unique geography and many densely-populated cities along coastal lines. Its peninsular shape and key islands have limited hurricane evacuation options.

Data employed in this research include the road network downloaded from the Florida Department of Transportation (FDOT; Data access: http://www.fdot.gov/statistics/hwydata/default.shtm). As interstate and major highways are most commonly used by evacuees (Dow and Cutter, 2002), this study only considers the national and state highway network. The network data contain the free-flow travel time and number of lanes of each road segment. The number of lanes of a road allows the estimation of its carrying capacity. As most people evacuate by automobiles, the number of vehicles would be a more reasonable representation of evacuation demand than the number of evacuees, and such data—the number of vehicles owned by each household (in 2016)—is acquired from the United States Census Bureau at the county subdivision (sub-county) geographic level (Data access: https://www.census.gov/geographies/mapping-files/time-series/geo/tiger-data.2016.html). In total, there are 316 sub-counties in Florida.

## 3. Methodology

As stated previously, estimations of evacuation demand and evacuees' route choices are two essential inputs in measuring road accessibility during a hurricane evacuation. In essence, both factors are significantly affected by a hurricane's forecasted track. Specifically, a threatening hurricane with the cone of uncertainty covering the whole of peninsula Florida, such as Irma, would leave residents no choice but to evacuate out of the state. Here assumes the evacuation as a merging behavior, where residents are heading towards the same direction using the shortest routes available. In Florida, a state which has a peninsula shape, most evacuees would travel northwards into adjunct states. The proposed method first estimates the evacuation demand in each sub-county by considering the hurricane's wind radius and track. The closest facility analysis is then employed to model evacuees' route choices towards the predefined evacuation destinations. Next, the crowdedness level (*PCI*) of each road is calculated by combining the estimated evacuation demand and route choices. Finally, the road accessibility of each sub-county can be derived from cumulating *PCI* values of participating roads leading the sub-county evacuees to their preferred destinations. See Figure 1 for the method workflow.

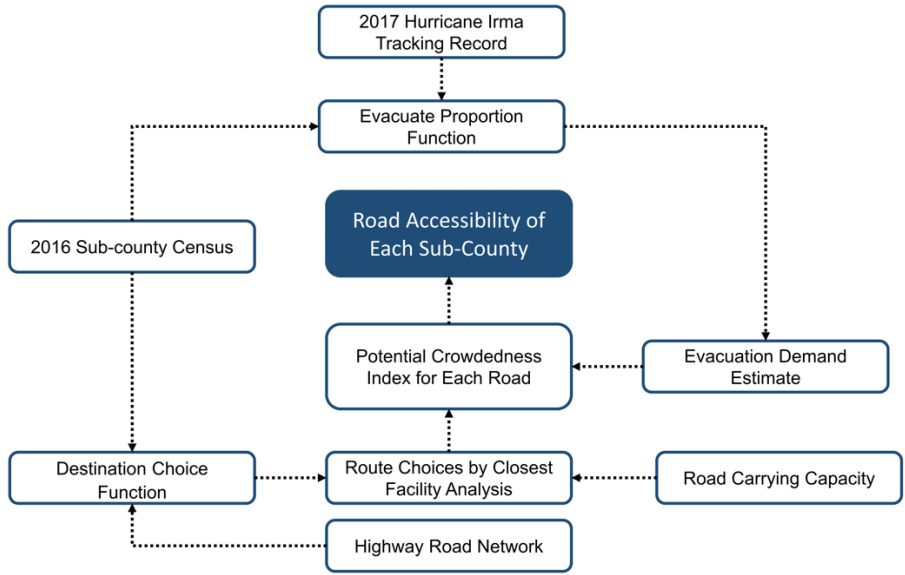

**Figure 1** Workflow of the proposed method

### *3.1 Evacuation Demand Estimation*

Different from the topological-based approach that assumes a constant network demand, the proposed method considers dynamic evacuation flow through the track of the hurricane. This is where the hurricane characteristics, such as wind speed and track, are taken into consideration. Specifically, this is achieved by integrating the behavioral response curves—implemented by various distance decay functions—in evacuation demand modeling. In general, these distance decay functions operate in a way that the estimated number of evacuees in an area is negatively related to the distance to the hurricane from that area. Put it in another way, residents' willingness to evacuate is related to the inverse distance from the hurricane (Lindell et al., 2005; Fu & Wilmot, 2006; Fu et al., 2007). Various types of decay functions have been explored, including, among others, natural (a decay coefficient of *e*), square (a decay coefficient of 2), and cubic (a decay coefficient of 3) (Fu & Wilmot, 2006; Fu et al., 2007). In implementation, a minimum distance threshold to the hurricane is usually applied. This is because it would be too risky for residents to evacuate when they are in close proximity to the hurricane. For example, Fu & Wilmot (2006) found that a distance threshold of 94 miles from the hurricane center to a household was appropriate for their case study, and evacuation would not be a rational choice for households within such a threshold. Given that the wind speed of a tropical cyclone is at least 34 knots (equivalent to 39 mph) and the employed hurricane tracking data are recorded in an interval of 5 knots, this study sets a wind speed of 35 knots (40 mph) as the distance threshold. By employing Hurricane Irma tracking data from the fourth version of International Best Track Archive for Climate Stewardship (IBTrACS; Knapp et al., 2010) where a tropical cyclone's status is reported every six hours, this study defines the distance from Hurricane Irma as the straight-line distance from each sub-county centroid to the 40-mph wind radius catchment of Hurricane Irma with a frequency of every six-hour from 48 hours before it made landfall (09-08-2017, 00:00) until it covered the entire state (09-11-2017, 12:00). These distances are then normalized and referred to as Distance Proportion (*DP*):

$$DP = \frac{D_c}{D_{Max}} \quad (1)$$

where $D_c$ is the straight-line distance from a sub-county centroid to the perimeter of the 40-mph wind radius of Hurricane Irma at a particular time, while $D_{Max}$ refers to the longest distance among all sub-county centroids (i.e., the distance between Escambia County, FL and the 40-mph wind radius catchment of the hurricane 48 hours before landfall—see Figure 2) and here it is used for standardization purposes. Once a sub-county centroid falls in the hurricane's 40-mph wind radius catchment area, residents in that sub-county are assumed not to be evacuating anymore (see Figure 2).

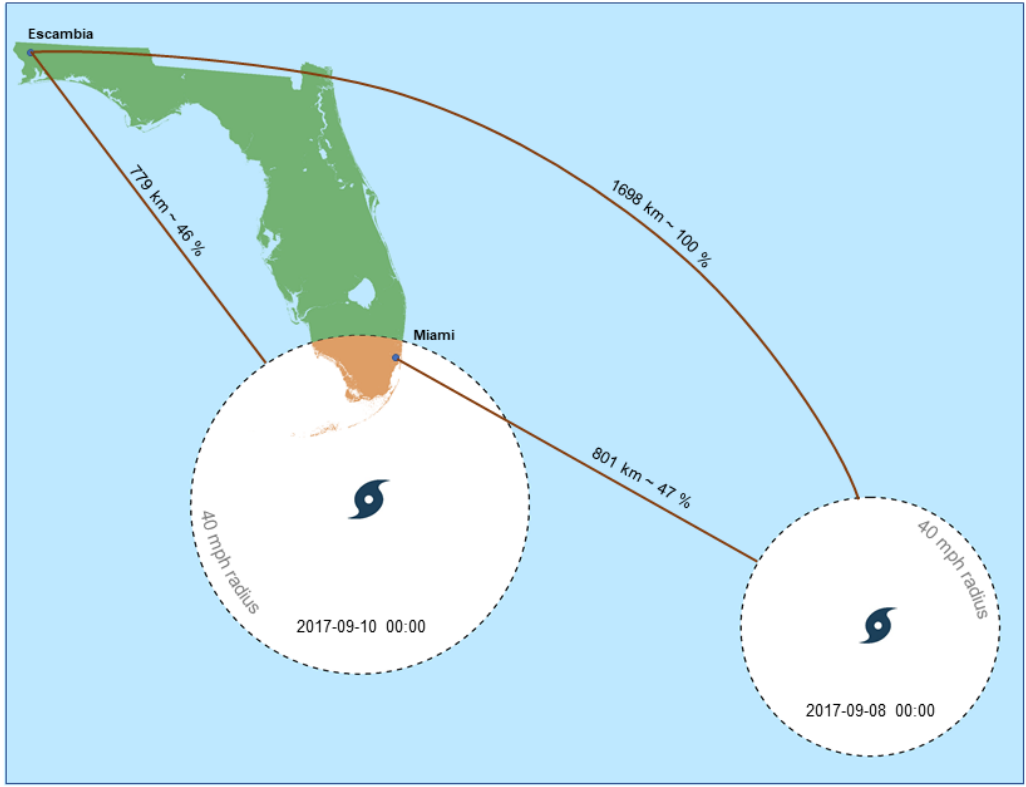

**Figure 2** Calculation of Distance Proportion of a sub-county [the longest line on top is curved for better visualization]

The nearer a sub-county is from the 40-mile wind radius catchment area of the hurricane, the greater the proportion of its residents are to evacuate. In this study, the natural decay function (neutral response curve) is employed. Other decay functions such as the square (aggressive response curve) and cubic (conservative response curve) are also examined in subsequent sections for a sensitivity analysis. The proportion of evacuation population (again, represented by number of household vehicles) in a sub-county is termed as Evacuate Proportion (*EP*):

$$EP = 1 - DP^{\alpha^{-1}} \quad (2)$$

where *DP* refers to the standardized distance from a sub-county to the 40-mph wind radius catchment of a hurricane (see Equation 1), $\alpha$ is the decay coefficient and is set to $e$, 2, 3 for natural, square, and cubic decay functions, respectively. The Evacuation Demand (*ED*) in each six-hour period is then defined as:

$$ED_{a+6} = ED_a \times (1 - EP_a) \times (1 - EP_{a+6}), \quad for\ (a = 0, 6, 12, 18, 24 \dots) \quad (3)$$

where $a$ is the recording period of Hurricane Irma.

*3.2 Route Choice and Network Assignment*

Assessing road network accessibility during a hurricane evacuation requires understandings of road usage and traffic flow distributions. This study uses the closest facility analysis in GIS to achieve this goal. Based on the defined road network, the shortest network paths for all pairs of origins and destinations (OD) are calculated. The origins are set using sub-county centroids. This study assumes evacuation vehicles flow towards the northern border of Florida and then travel out of the state. Four places along the northern boundary of Florida (Jennings, Berrydale, Havana, and Callahan-Hilliard) are selected as destinations because they are at junctions of major interstate highways. The output of the closest facility analysis gives visualized route choices from each origin to the destinations. The shortest paths of these OD pairs are plotted in Figure 3 where origins are displayed in dots and destinations in cross marks.

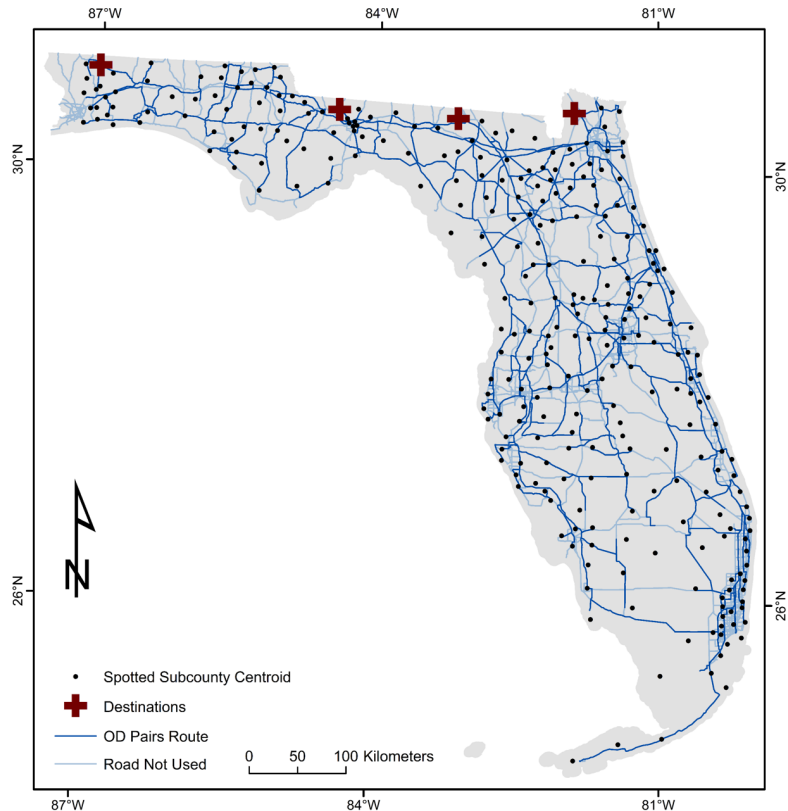

**Figure 3** Visualization of route choices solved by the closest facility analysis

Note that the four destinations are not evenly assigned for sub-counties, since the destination closer to an origin centroid is more likely to be chosen. Therefore, this study measures the network distance between an origin and each of the four destinations and uses cubic decay function as an estimation of the Destination Choices Weight ($DCW$). The cubic decay function is employed here in order to emphasize the distance decay effect on evacuees' destination choices and distinguish population flows among the four destinations:

$$DCW(d_{ij}) = \frac{(\sum_{k=1}^{4} d_{ik}/d_{ij})^3}{\sum_{k=1}^{4}(\sum_{k=1}^{4} d_{ik}/d_{ij})^3} \qquad \begin{cases} i \in \{1,2,3 \dots 312\} \\ j \in \{1,2,3,4\} \\ k \in \{1,2,3,4\} \end{cases} \qquad (4)$$

where $d_{ij}$ refers to the network distance between origin $i$ and destination $j$, and $k$ represents a destination site.

### 3.3 Potential Crowdedness Index (PCI)

The previous step assigns evacuation flow into the road network. For each road segment, the assigned traffic flow is then compared with its carrying capacity to derive the potential crowdedness level. For a road segment, congestion occurs when evacuation demand assigned to it exceeds its carrying capacity (Qian et al., 2006). On a regional scale, the ratio between road carrying capacity and demand, also known as the Bulk Lane Demand ($BLD$), could, therefore, express the level of crowdedness or congestion (Chen et al., 2012). It is expressed by:

$$BLD = \frac{P}{C} \quad (5)$$

where $P$ denotes the evacuation demand, which is the number of vehicles assigned to the road. Road carrying capacity, $C$, is represented by the number of lanes of that road. As some road segments may be traversed by multiple OD pairs, the $BLD$ is then adjusted to the Potential Crowdedness Index ($PCI$) for a more accurate representation of congestion level for each road segment:

$$PCI = \frac{\sum P}{C} \quad (6)$$

where the road demand becomes a summation of vehicles ($P$) traversing the same road segment.

### 3.4 Road Accessibility of Sub-Counties during a Hurricane Evacuation

The $PCI$ represents the crowdedness or congestion level of a road segment, while road accessibility depicts the collective ease of a sub-county's evacuees in reaching a destination through the road network. In other words, the two metrics measure opposing patterns at different scales—$PCI$ about the level of *difficulty* for evacuees to move along a *linear* feature while road accessibility about the level of *ease* to move for evacuees from an *areal* feature. Such a contrasting relationship is analogous to that between a neighborhood's health care accessibility measured by the two-step floating catchment area (2SFCA) method and a health care facility's crowdedness level measured by the i2SFCA (inverted 2SFCA) method (Wang, 2018). Therefore, road accessibility of a sub-county is measured by calculating the reciprocal of the sum of *PCI* values associated with the road segments on the path from this sub-county to the four designated destinations. This calculation step would, however, discount road accessibility for those remote sub-counties from any of the predefined destinations, such as those in the very south, simply because of the greater number of road segments involved. Therefore, the road accessibility for each sub-county $i$ is finally normalized by the total length of trips from sub-county $i$ to all the four destinations, and it is formulated as:

$$A(i) = \frac{1}{\sum_{j=1}^{4}[DCW(d_{ij}) \times \sum PCI(d_{ij})/d_{ij}]}, \qquad \begin{cases} i \in \{1,2,3 \dots 316\} \\ j \in \{1,2,3,4\} \end{cases} \quad (7)$$

and a sub-county with a higher value of *A* indicates that sub-county having better road accessibility. To avoid biases that could occur for sub-counties in close proximity to the four destinations, only sub-counties south to 30° N are assessed here.

## 4. Results

In order to more fully understand the road accessibility pattern, the entire lifespan of Hurricane Irma is separated into pre-landfall and post-landfall time groups in this research. The pre-landfall period (09-08-2017, 00:00 to 09-10-2017, 00:00) is defined as the first 48 hours before the hurricane made landfall, while the post-landfall period (09-10-2017, 06:00 to 09-11-2017, 12:00) is defined as the time until the end period. The proposed method is then applied to calculate road accessibility of sub-counties for both pre- and post-landfall time groups.

a

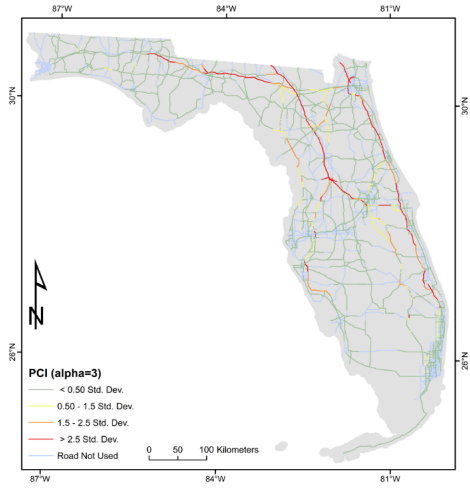 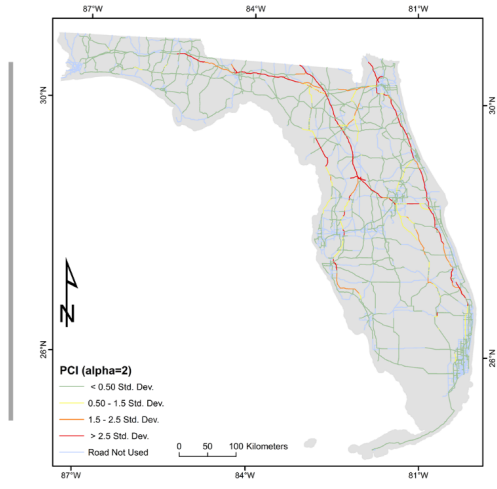 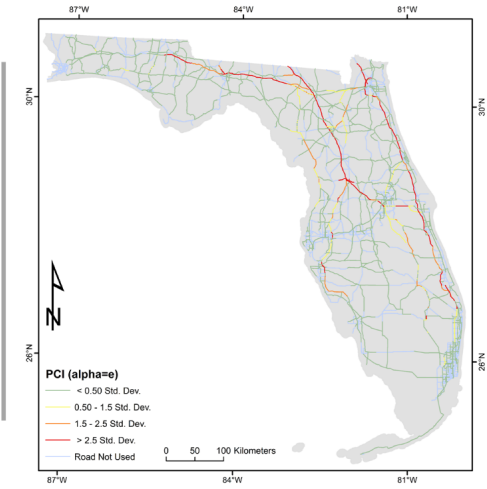

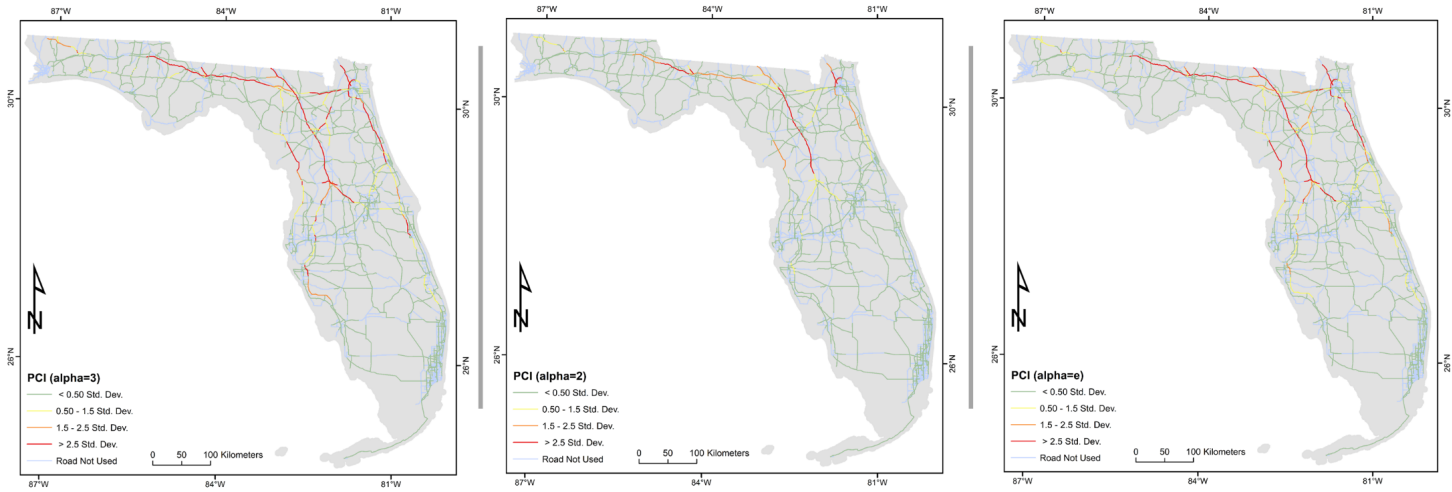

Figure 4 Potential crowdedness index of road segments associated with three behavioral response curves between: **a**. pre-landfall and **b**. post-landfall

Figure 4 shows the *PCI* values associated with each road segment between pre- and post-landfall periods. As shown in both time groups, highway segments serving populous sub-counties expectedly experience severe level of crowdedness. For example, I-75 has the highest crowdedness value because of its close proximity to large cities such as Tampa and Orlando.

The southern part of Highway I-95 has a greater crowdedness level at the pre-landfall stage than post-landfall. This is because the majority of evacuees are from the southeastern coastal area at the pre-landfall stage. During the post-landfall period, inland highways and the northern part of I-95 are under greater pressure compared to those from the southeastern coastal areas. Evacuees at that stage are mainly from inland and northern regions.

Figure 5 shows the road accessibility of sub-counties between pre- and post-landfall periods. For both pre- and post-landfall stages, sub-counties on the northeastern coast of the peninsula are reported to have lower road accessibility than the western coast because the roads in this area are also largely used by evacuees from the south, such as the Miami region. The concentration of poor road accessibility partially shifts from the southeastern coast to the central area after the hurricane made landfall. Accumulation of evacuation traffic from the south triggered high traffic stress near the evacuation destinations. The result is consistent with the pattern of the road potential crowdedness because places next to road segments of high usage are expected to experience higher delays than usual.

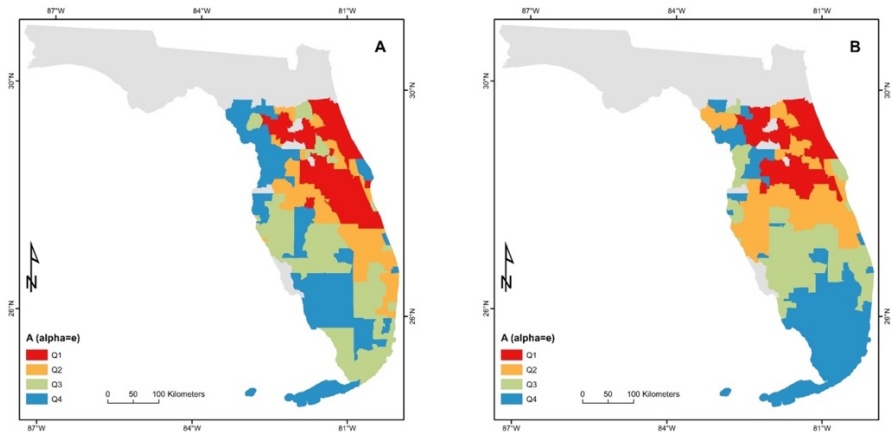

**Figure 5** Sub-county road accessibility between: **a**. pre-landfall and **b**. post-landfall

Some inland sub-counties in close proximity to major highways also experienced lower than average road accessibility in spite of fewer evacuees. This is because the great evacuation traffic from the southern coastal area percolates to the inland road network. Therefore, less populous areas may also experience poor road accessibility.

## 5. Discussion
### 5.1 Sensitivity Analysis of Behavioral Response Curves

Out-of-state evacuations in Florida due to a hurricane's passage will lead to severe congestion on major highways, such as the reported 1,236% increase in the observed traffic volume on northbound I-75 during Hurricane Irma (Florida Department of Transportation, 2018). This reported fact is consistent with our finding that I-75 will be in a massive crowded condition reflected by the high *PCI* value.

The spatial variation of potential road congestion is determined by the total number of vehicles, road carrying capacity, distance to the hurricane, and direction of the evacuation traffic. However, it is dependent on the choice of behavioral response curves that are reflective of evacuees' risk perceptions. The choice of the decay function could significantly affect the results (Baker, 1991). Therefore, a sensitivity analysis is performed to examine how different choices of decay functions could have affected the estimation of road demand and crowdedness over time. The increasing cumulative percentages of estimated vehicles associated with the three selected decay functions slow down after the hurricane made landfall (Figure 6a), which is consistent with the departure time pattern concluded by Hasan et al. (2013) from an empirical case study of Hurricane Ivan.

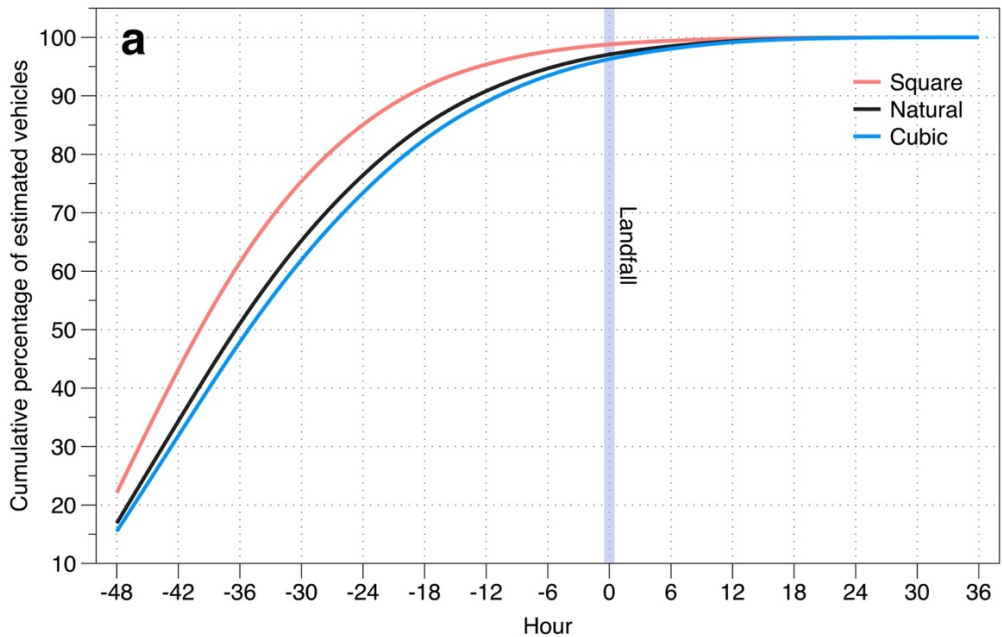

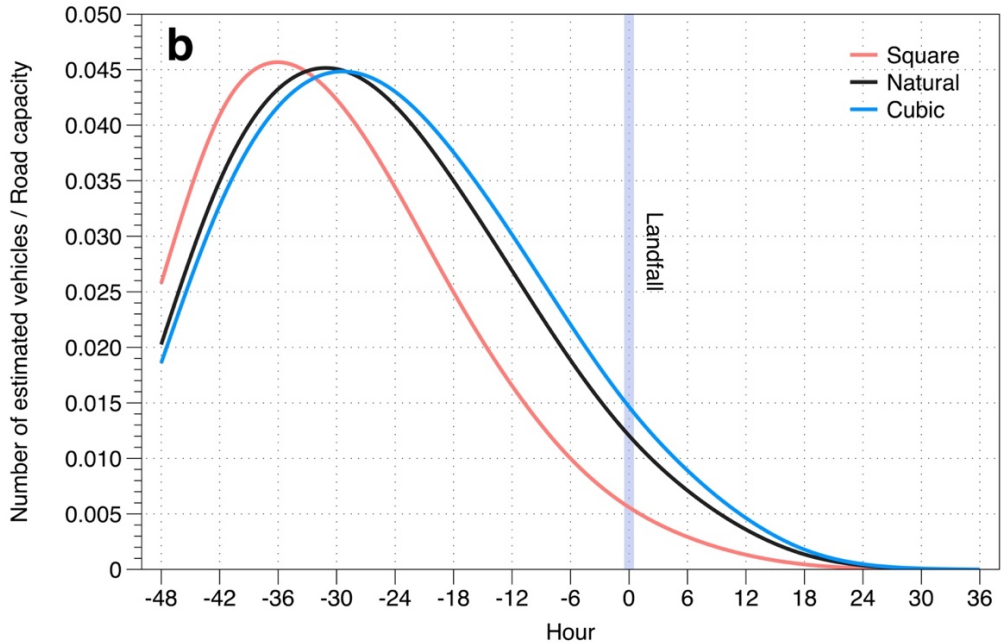

**Figure 6** Sensitivity analysis using three decay functions: **a**. cumulative percentage of estimated vehicles on roads and **b**. level of congestion for the road network

Among all three employed decay functions, the square decay function shows the most rapid increase in terms of the time step due to more aggressive settings ($\alpha = 2$) of residents' reactions to risks. As a consequence, more evacuees were observed at the pre-landfall stage compared to the other two functions. The increased number of evacuees associated with the square decay function may overload the road network, however. It is found that the overall level of congestion for the road network—estimated by the total number of estimated vehicles on roads divided by the total road capacity—reaches the peak relatively earlier than the other two functions (Figure 6b), causing evacuation to move slowly at the early stage but allowing more residents to evacuate before Hurricane Irma made landfall. Is it worthwhile to risk overloading the road network in order for more people to leave when they can? Table 1 shows a detailed comparison of traffic conditions before and after the hurricane made landfall across the three decay functions. It is clear that the switch from a natural decay to square decay function, for example, gives rise to a 3.3% gain ((43.1 – 41.7) / 41.7) in predicted number of vehicles at the pre-landfall stage. This, however, leads to only an overall 2.4% downgrade in network congestion. This may indicate the effectiveness of adopting such a rapid response curve in emergency management and analysis. In addition, the choice of the decay function has varying effects between the two stages. Looking at the natural and square decay function again, at the post-landfall stage, a more rapid response from residents leads to, however, a 59.3% decrease in the number of vehicles on roads and a 58.3% improvement in road congestion level. A closer view at the three major highways (I-75, I-95, and I-10) in northeastern Florida (Figure 7) shows that, compared to the natural decay function, the square decay function gives rise to evident reductions (about 50%) in the level of congestion in these segments.

Table 1 Estimated evacuation demand and network crowdedness across three decay functions

| Decay Function Coefficient ($\alpha$) | Total Number of Vehicles ($x10^5$) | | Network Potential Crowdedness ($x10^7$) | |
| --- | --- | --- | --- | --- |
| | Pre-landfall | Post-landfall | Pre-landfall | Post-landfall |
| $e$ | 41.7 | 2.7 | 148.7 | 5.3 |
| $2$ | 43.1 | 1.1 | 152.2 | 2.2 |
| $3$ | 40.9 | 3.3 | 146.9 | 6.7 |

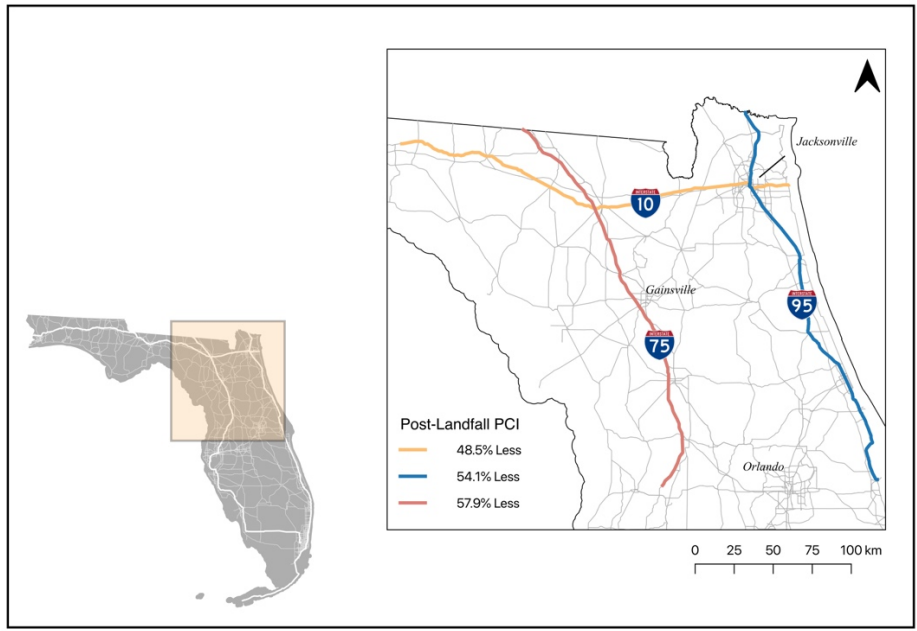

Figure 7 Percent reductions in *PCI* values in I-75, I-95, and I-10 in northeastern Florida for the square decay function relative to the natural decay function

*5.2 Implications on Hurricane Hazards Mitigation*

Clearly, the choice of behavioral response curves (i.e., distance decay functions) can affect the estimation of road potential crowdedness, which will subsequently have an effect on sub-county residents' road accessibility. The results indicate that, compared to the natural decay function, the square decay function results in an over five-percent decrease in road accessibility on the northwestern coast of the peninsula before Hurricane Irma made landfall (Figure 8).

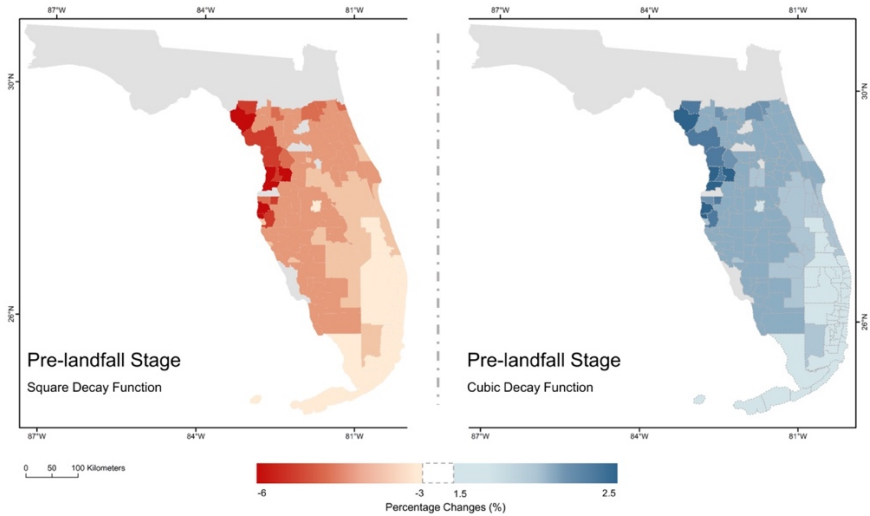

**Figure 8** Percentage changes (relative to the natural decay function) of sub-counties' road accessibility level before Hurricane Irma made landfall

Residents in sub-counties along the northwestern coast are farthest to the hurricane's destructive wind radius and are less likely to evacuate. However, the square decay function assumes residents being highly sensitive to the distance to the hurricane and therefore results in a greater decrease in road accessibility in the area. Comparatively, the cubic decay function (a more conservative evacuation behavior) results in an over two percent accessibility increase along the northwestern coastal area. For the post-landfall period (Figure 9), sub-counties along the southeastern coast of the peninsula are predicted to experience greater differences than other areas. For the conservative cubic decay function case, due to fewer vehicles participating in the pre-landfall evacuation period, in the post-landfall period, an average of a 25-percent decrease in road accessibility across the peninsula is observed, with the southeastern coastal area having the lowest road accessibility. In contrast, a higher evacuation rate associated with the square decay function before the hurricane made landfall results in as much as a 60-percent increase in road accessibility in the eastern coastal area after the hurricane made landfall.

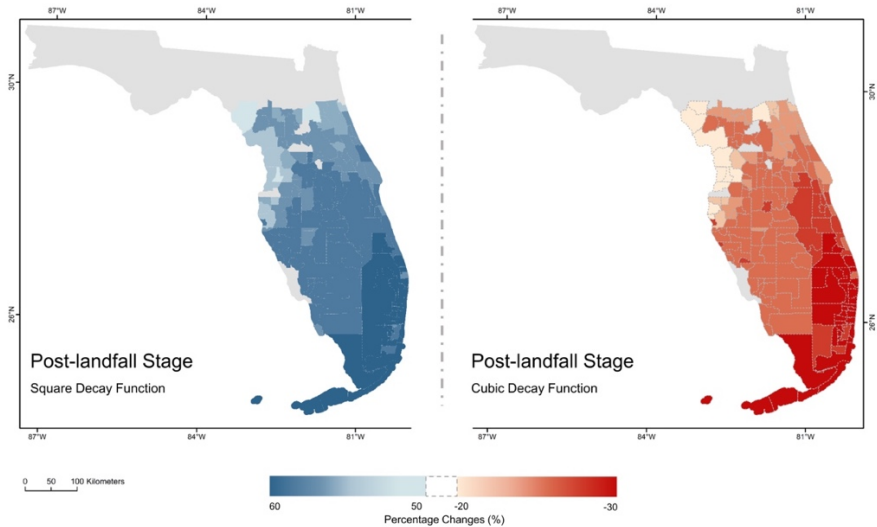

**Figure 9** Percentage changes (relative to the natural decay function) of sub-counties' road accessibility level after Hurricane Irma made landfall

The above analyses demonstrate the impact of different response curves from residents on the overall evacuation performance and road accessibility. Clearly, a more rapid response from participants, such as $a = 2$ in this study, could lead to more people evacuating before a hurricane makes landfall and thus may help reduce life losses. Such knowledge may assist emergency management officers and researchers in determining more appropriate parameters for their predictive models. Several studies have shown that factors such as warning messages from official emergency agencies (Sorensen, 2000; Dow & Cutter, 2000), perceptions of geophysical hazards (Meyer et al., 2018), and social behaviors (Dash & Gladwin, 2007; Collins et al., 2017; 2018) could influence residents' behavioral response curves. However, emergency management officials may choose not to promote a rapid response curve all the time so as to minimize the occurrence of false alarms (Dow & Cutter, 1998). And if they choose to adopt such a method to encourage people to evacuate early, additional operational and management strategies such as a staged evacuation plan should be executed to minimize the number of shadow evacuees. In any case, it is beneficial for decision-makers to be aware of the relationship between participants' behavioral response curves and the overall evacuation and network performance.

Evacuation in a geographically restricted area, such as a peninsula, is quite different from that in continental coastal areas. In addition, the nature of a hurricane is also unique among all other hazards; it usually has a large impact area and follows a particular direction of movement. For an intense hurricane forcing an out-of-state evacuation from Florida, residents could only head northwards in generally the same direction. It is therefore worth paying close attention to an intense hurricane that may impact a peninsula as it may trigger the bottle-neck phenomenon.

## 6. Conclusions

In this paper, a methodological framework for measuring the road network accessibility of residents during a hurricane evacuation is proposed. Distinguished from existing studies in this scope, the proposed method takes into account both dynamic evacuation demand and the characteristics of a hurricane. In an interval of every six-hours, the method first estimates the evacuation demand in terms of number of vehicles per household in each sub-county by considering the hurricane's wind radius and track. The closest facility analysis is then employed to model evacuees' route choices towards the predefined evacuation destinations. The potential crowdedness index ($PCI$), a metric capturing the level of crowdedness of each road segment, is computed afterward by coupling the estimated evacuation demand and route choices. Finally, the road accessibility of each sub-county is measured based on $PCI$ values of corresponding roads connecting evacuees from that sub-county to the designated destinations. The method is applied to the entire state of Florida during Hurricane Irma in September 2017 as a case study, and a sensitivity analysis is performed to understand the impacts of evacuee's behavioral response curves—represented by a variety of distance decay functions—on road accessibility measurements. Results show that I-75 and I-95 northbound have a high level of crowdedness, and sub-counties along the northbound I-95 have the lowest road accessibility. This implies a bottle-neck phenomenon that results from a large-scale evacuation order, particularly on a peninsula like Florida. The proposed method as well as the findings can help emergency agencies identify potentially congested road segments and highlight vulnerable neighborhoods of poor road access, and hence can shed light on evacuation management and transportation planning. The derived spatial maps of the road congestion level can be used by transportation and emergency management researchers and practitioners for further analysis such as network performance analysis to provide operational and managerial insights. Applications in other fields such as understanding the impacts of urban flooding on neighborhoods' health care access (Balomenos et al., 2019) can benefit from the results as well. In addition, the sensitivity analysis suggests that the choice of an aggressive behavioral response curve might be worthwhile to be considered in the evacuation demand modeling and decision-making process—and certainly should be implemented along with other strategies such as a staged evacuation plan—as it could lead to more people leaving vulnerable areas before the arrival of a hurricane in spite of putting a burden on the road network. Compared to the degradation in the network congestion level, the population gain in terms of evacuees is found to be more significant.

There are some limitations that merit discussion. First, the study specifically examines Hurricane Irma where evacuees, in general, traveled northwards to evacuate. Such a directional preference for evacuees from a large region such as a state may not work well for other areas that do not have a peninsula shape. Second, the estimation of evacuation demand through road network is dependent on the choice of the evacuee's behavioral response curves concerning the distance to the hurricane's catchment area defined by its wind speed. Some other factors, such as demographic (race and gender) and people's risk perceptions, especially the perception of the safety on their own homes, as suggested in Meyer et al. (2018), may also play a major role in affecting people's evacuation decision-making and need to be examined in future studies. Third, a more systematic way to select evacuation destinations could be beneficial. One way to do so is to look at historical traffic data, if available, and infer the most "popular" destinations. Fourth, the assumption in traffic assignment that evacuees use the shortest path for evacuation may not be realistic, as there is an increasing tendency for evacuees to follow route choices suggested by a GPS navigation system (e.g., to avoid highways). However, a recent study that examined evacuee behavior during Hurricane Irma using collected survey data found that the actual impact of such GPS navigation systems on users' final routing choices were minimal (Wong et al., 2018). Nevertheless, the assumption of evacuees being utility-maximizers and the implementation based on travel distance without accounting for real-time traffic can still bring about relatively meaningful estimates on traffic patterns (Hu and Downs, 2019). Fifth, the assumption that residents who fall within the 40-mph wind radius catchment of a hurricane would no longer evacuate may underestimate evacuation demand by omitting post-storm evacuation due to issues such as power outages and disease spread. Sixth, the proposed methods would benefit from the consideration of population dynamics in a more frequent time interval, such as hourly population distribution. Such information could be obtained by analyzing mobile phone positioning data. Similarly, the road network may undergo significant changes during an evacuation—some roads might be closed and counterflow might be established (Dow & Cutter, 2002; Balomenos et al., 2019). Future steps of this study should include scenario-based network analysis to investigate future network performances coupled with projected population, hurricane, and road network availability. Also, future research will seek to acquire Hurricane Irma's traffic data, such as traffic counts at stations along major roads, as it would be a valuable resource to validate the results. Finally, in order to provide more practical implications with decision-makers, some further analysis that looks into minimizing traffic congestion by finding the optimal order of evacuation by sub-counties or even smaller zones is warranted.

## Acknowledgements
Dr. Yujie Hu would like to acknowledge the support by a grant from the Center for Transportation Equity, Decisions and Dollars (CTEDD) funded by U.S. Department of Transportation Research and Innovative Technology Administration (OST-R). We also thank the anonymous referees for their valuable comments.